# An empirical study of LoRA-based fine-tuning of large language models for automated test case generation

Milad Moradi[*]

AI Research Lab, Tricentis, Vienna, Austria

`m.moradi-vastegani@tricentis.com`

Ke Yan

AI Research Lab, Tricentis, Sydney, Australia

`k.yan@tricentis.com`

David Colwell

AI Research Lab, Tricentis, Sydney, Australia

`d.colwell@tricentis.com`

Rhona Asgari

AI Research Lab, Tricentis, Vienna, Austria

`r.asgari@tricentis.com`

---

[*] Corresponding author. **Postal address:** Tricentis GmbH, Leonard-Bernstein-Straße 10, 1220 Vienna, Austria.

## Abstract

Automated test case generation from natural language requirements remains a challenging problem in software engineering due to the ambiguity of requirements and the need to produce structured, executable test artifacts. Recent advances in Large Language Models (LLMs) have shown promise in addressing this task; however, their effectiveness depends on task-specific adaptation and efficient fine-tuning strategies. In this paper, we present a comprehensive empirical study on the use of parameter-efficient fine-tuning, specifically Low-Rank Adaptation (LoRA), for requirement-based test case and test step generation. We evaluate multiple LLM families, including open-source models (DeepSeek-1.8B, Llama-3.1-8B, and Ministral-8B) and proprietary models (GPT-4.1, GPT-4.1-mini, and GPT-4.1-nano), under a unified experimental pipeline. The study systematically explores the impact of key LoRA hyperparameters, including rank, scaling factor, and dropout, on downstream performance. To enable scalable and consistent evaluation, we propose an automated evaluation framework based on GPT-4o, which assesses generated test cases across nine quality dimensions, such as semantic similarity, information coverage, structural correctness, and hallucination. Experimental results demonstrate that LoRA-based fine-tuning significantly improves the performance of all open-source models, with Ministral-8B achieving the best results among them. Furthermore, we show that a fine-tuned 8B open-source model can achieve performance comparable to pre-fine-tuned GPT-4.1 models, highlighting the effectiveness of parameter-efficient adaptation. While GPT-4.1 models achieve the highest overall performance, the performance gap between proprietary and open-source models is substantially reduced after fine-tuning. These findings provide important insights into model selection, fine-tuning strategies, and evaluation methods for automated test generation. In particular, they demonstrate that cost-efficient, locally deployable open-source models can serve as viable alternatives to proprietary systems when combined with well-designed fine-tuning approaches, offering practical benefits in terms of scalability, customization, and data privacy.

# 1. Introduction
## 1.1. Background and motivation

Software testing is a critical yet resource-intensive phase of the software development lifecycle [1]. Among testing activities, test case generation is particularly challenging, as it requires translating often informal, ambiguous, and incomplete requirements into precise, structured, and executable test artifacts [2, 3]. Despite decades of research and industrial practice, generating high-quality test cases that are correct, complete, and maintainable remains a largely manual task, demanding significant domain expertise and human effort. As software systems continue to grow in size, complexity, and release frequency, the limitations of manual test case design have become increasingly evident, motivating the need for more effective automation solutions [3].

Automated test case generation from requirements faces several inherent challenges. First, software requirements are commonly expressed in natural language, which is prone to ambiguity, inconsistency, and varying levels of abstraction. Accurately interpreting requirement intent and translating it into concrete test cases and test steps is non-trivial. Second, test cases are structured artifacts that must follow specific conventions, including test case names, types, descriptions, and ordered test steps with clear preconditions and expected outcomes. Generating such structured outputs requires not only language understanding but also adherence to testing best practices. Third, practical test case generation must balance coverage, correctness, and readability, ensuring that generated tests are both technically valid and useful for human testers and automation frameworks. These challenges have limited the effectiveness of traditional automation techniques [4-6].

Early approaches to automated test case generation largely relied on rule-based systems, templates, and heuristic methods. While effective in narrowly defined domains, rule-based techniques suffer from poor scalability and maintainability. They require extensive manual rule engineering and struggle to generalize across projects, domains, or requirement styles. Even small changes in requirement phrasing often necessitate significant updates to the underlying rules [7, 8].

Traditional machine learning approaches, including classical NLP models and supervised learning techniques, have attempted to alleviate some of these issues. However, these methods typically depend on feature engineering, domain-specific representations, and limited contextual understanding [9]. As a result, they often fail to capture long-range dependencies in requirements or generate coherent, multi-step test procedures. Moreover, conventional ML models generally lack the expressive capacity required to produce rich, structured outputs such as detailed test steps, making them insufficient for end-to-end test case generation in real-world settings [10].

Recent advances in Large Language Models (LLMs) have fundamentally changed the landscape of natural language processing and have opened new opportunities for software engineering automation. Trained on massive corpora of natural language and code, LLMs demonstrate strong capabilities in understanding context, reasoning over complex inputs, and generating structured, human-readable outputs. Consequently, they have been successfully applied to a wide range of software engineering tasks, including code generation, bug detection, documentation, requirements analysis, and software testing [11, 12].

In the context of test automation, LLMs offer the potential to directly transform natural language requirements into complete test cases, including descriptive metadata and executable test steps [13, 14]. However, off-the-shelf LLMs are typically trained as general-purpose models and may not perform optimally on specialized downstream tasks such as requirement-based test case generation without further adaptation. This has motivated growing interest in fine-tuning LLMs for software engineering tasks, particularly using parameter-efficient methods that reduce computational cost while preserving performance [15]. Understanding how different models and fine-tuning strategies perform in this domain is therefore essential for advancing practical, scalable solutions to automated test case generation.

## 1.2. Problem statement

The growing adoption of agile and DevOps practices has intensified the demand for rapid and reliable software testing, highlighting the need for automated generation of test cases directly from software requirements [16]. In many industrial settings, requirements are specified in natural language and serve as the primary source of truth for defining system behavior. Transforming these requirements into high-quality test cases (comprising a test case name, test case type, test case description, and a sequence of well-defined test steps) remains a largely manual and error-prone process [17]. This manual effort not only increases development cost but also introduces inconsistencies between requirements and test artifacts, particularly in fast-evolving systems.

From a technical perspective, automated requirement-based test case generation poses a complex text-to-structured-artifact transformation problem. Given a requirement name and its textual description as input, an automated system must correctly interpret the intended functionality, identify relevant test scenarios, and generate structured and logically ordered test steps that align with software testing best practices [18]. The generated test cases must be sufficiently detailed to be actionable, while remaining concise and readable for both human testers and automated testing frameworks. Achieving this level of quality and consistency across diverse requirement formulations is a significant challenge.

LLMs have recently emerged as promising candidates for addressing this problem due to their strong natural language understanding and text generation capabilities [19]. However, off-the-shelf LLMs are not optimized for domain-specific tasks such as test case generation. Without adaptation, they may produce incomplete test steps, irrelevant scenarios, or outputs that do not conform to expected testing structures. Full fine-tuning of large models can improve task performance but is often computationally expensive, memory-intensive, and impractical for many organizations, especially when dealing with models containing billions of parameters [20].

This leads to a critical need for efficient and high-quality fine-tuning strategies that can adapt LLMs to requirement-based test case generation without incurring prohibitive costs. Parameter-efficient fine-tuning methods, such as Low-Rank Adaptation (LoRA), offer a promising solution by updating only a small subset of model parameters while keeping the base model frozen [21]. However, the effectiveness of such approaches depends heavily on the choice of fine-tuning configuration and hyperparameters, and their impact varies across different model architectures.

Therefore, the central problem addressed in this work is twofold:

1. how to effectively fine-tune LLMs to automatically generate high-quality test cases and test steps from natural language requirements, and
2. how to achieve this adaptation efficiently, balancing performance, computational cost, and model size.

Addressing this problem requires a systematic investigation of fine-tuning strategies across different LLM families, along with robust evaluation methods to assess the quality of generated test artifacts.

### 1.3. Research objectives and contributions

The overarching objective of this research is to systematically investigate the effectiveness of parameter-efficient fine-tuning of LLMs for automated test case and test step generation from natural language requirements. While recent studies have demonstrated the promise of LLMs in software engineering tasks [12, 14], there is still limited empirical evidence on how different model families and fine-tuning strategies perform for requirement-based test automation, particularly under realistic resource constraints. This work aims to close this gap through a large-scale, methodical experimental study.

To achieve this objective, we conduct a comprehensive empirical evaluation of LoRA-based fine-tuning applied to multiple state-of-the-art LLMs. Rather than relying on anecdotal or single-configuration results, our study explores a wide range of fine-tuning settings and analyzes their impact on the quality of generated test cases. This allows us to better understand how parameter-efficient adaptation influences model behavior in the context of structured test artifact generation.

A key contribution of this paper is a comparative evaluation between open-source and proprietary models. Specifically, we evaluate several open-source LLMs at the 8B parameter scale and compare their fine-tuned performance against fine-tuned GPT-4 models provided via Microsoft Azure. This comparison provides practical insights into the trade-offs between model size, openness, cost, and achievable performance, which are critical considerations for both researchers and practitioners.

In addition, this work presents a systematic exploration of LoRA hyperparameters, including rank, scaling factors, and other configuration choices. By analyzing how these parameters affect downstream performance, we demonstrate that LoRA configuration is not merely an implementation detail but a decisive factor in achieving high-quality results. Our findings offer concrete guidance for practitioners seeking to fine-tune LLMs efficiently for similar tasks.

To support scalable and consistent evaluation, we introduce an automated evaluation framework based on GPT-4o, which assesses generated test cases across multiple quality dimensions, such as requirement coverage, correctness, completeness of test steps, and clarity. This framework enables large-scale evaluation without the prohibitive cost of exhaustive human assessment and provides a reproducible mechanism for comparing different models and fine-tuning setups.

Finally, one of the most significant findings of this study is the empirical evidence that a fine-tuned 8B open-source model can achieve performance comparable to that of pre-fine-tuned GPT-4 models on the task of test case generation. This result highlights the potential of smaller, open-source models as viable and cost-effective alternatives to large proprietary models when combined with well-designed parameter-efficient fine-tuning strategies.

In summary, the main contributions of this paper are:

- A large-scale empirical study of LoRA-based fine-tuning for automated test case and test step generation.
- A detailed comparison of fine-tuned open-source 8B models and fine-tuned GPT-4 models.

- A systematic analysis of the impact of LoRA hyperparameters on downstream performance.
- An automated, LLM-based evaluation framework for assessing test case quality.
- Empirical evidence demonstrating performance parity between fine-tuned 8B open-source models and pre-fine-tuned GPT-4 models.

Together, these contributions advance the state of the art in LLM-based software test automation and provide actionable insights for both research and industrial adoption.

The rest of the paper is organized as follows. Section 2 reviews the background and related work on automated test case generation, large language models, and parameter-efficient fine-tuning. Section 3 defines the task formulation, dataset, and prompt design used in this study. Section 4 presents the methodology, including the experimental pipeline, model selection, and LoRA-based fine-tuning setup. Section 5 introduces the automated evaluation framework, while Section 6 reports and discusses the experimental results. Finally, Section 7 concludes the paper and outlines key findings and implications.

## 2. Background and related work

### 2.1. Automated test case generation

Automated test case generation has long been an active research area in software engineering, with requirement-based test generation being one of its most prominent directions [5, 22, 23]. In this setting, test cases are derived directly from software requirements to ensure traceability between specified behavior and validation artifacts. Early work in this area focused on leveraging structured or semi-structured requirements, such as use cases, UML models, or formal specifications, to systematically generate test scenarios and test steps [24, 25]. While these approaches provide strong alignment between requirements and tests, their applicability is often limited by the availability of well-structured requirement models and the effort required to maintain them as requirements evolve.

To overcome these limitations, researchers have increasingly explored Natural Language Processing (NLP) and Machine Learning (ML) techniques for test case generation from textual requirements. NLP-based methods typically rely on syntactic parsing, keyword extraction, and semantic role labeling to identify actions, conditions, and expected outcomes within requirement descriptions [19, 26]. Traditional ML approaches extend this by learning mappings between requirement features and test artifacts from historical data [27-29]. Although these techniques reduce manual effort compared to rule-based methods, they often depend on handcrafted features and domain-specific assumptions, which restrict their generalization capability. As a result, such approaches struggle to generate comprehensive and coherent multi-step test cases, motivating the adoption of more expressive models such as LLMs in recent research.

**2.2. Large language models in software engineering**

LLMs have recently gained significant attention in software engineering due to their strong capabilities in understanding and generating both natural language and source code [12, 30-32]. Trained on large-scale corpora that include programming languages, technical documentation, and natural language text, LLMs have been successfully applied to a variety of software engineering tasks. In code generation, they have demonstrated the ability to synthesize functions, complete code snippets, refactor existing code, and assist developers during implementation [33, 34]. These capabilities have led to increased productivity and have positioned LLMs as effective assistants throughout the software development lifecycle.

Beyond code generation, LLMs have also shown promise in software testing and requirements engineering [30]. In testing, prior work has explored the use of LLMs for generating unit tests, test inputs, test oracles, and test scripts, as well as for test prioritization and fault localization [12, 35]. In requirements engineering, LLMs have been applied to tasks such as requirement classification, ambiguity detection, requirement summarization, and traceability link generation [36, 37]. These studies highlight the potential of LLMs to bridge the gap between informal natural language artifacts and structured software engineering outputs. However, existing work often relies on pre-trained models with limited task-specific adaptation, leaving open questions regarding the role of fine-tuning and model configuration in achieving reliable and high-quality results for specialized tasks such as requirement-based test case generation.

**2.3. Fine-tuning large language models**

Fine-tuning is a common strategy for adapting pre-trained large language models to specific downstream tasks [38, 39]. Full fine-tuning updates all model parameters using task-specific data, often resulting in strong performance improvements. However, for modern LLMs with billions of parameters, full fine-tuning is computationally expensive, memory-intensive, and difficult to scale. It requires substantial hardware resources and can be impractical for many organizations, particularly when multiple task-specific models must be maintained. Additionally, full fine-tuning increases the risk of overfitting when training data is limited and complicates model deployment and versioning due to the need to store and manage large numbers of fine-tuned parameter sets [40].

To address these limitations, parameter-efficient fine-tuning (PEFT) methods have been proposed, which adapt LLMs by updating only a small subset of parameters while keeping the base model frozen [21]. PEFT techniques include adapter layers, prefix tuning, prompt tuning, and low-rank adaptation, all of which aim to reduce training cost while preserving most of the representational power of the pre-trained model [41]. By limiting the number of trainable parameters, PEFT enables faster training, lower memory consumption, and improved reproducibility, making it particularly suitable for industrial and large-scale experimental settings. Moreover, PEFT methods allow multiple task-specific adaptations to be stored and deployed efficiently on top of a shared base model.

Among PEFT techniques, Low-Rank Adaptation (LoRA) has gained widespread adoption due to its simplicity and effectiveness [42]. LoRA operates by decomposing weight updates into low-rank matrices that are injected into selected linear layers (commonly the attention and feed-forward projections) during fine-tuning [43]. Instead of learning full-rank weight updates, LoRA constrains the adaptation to a low-dimensional subspace, significantly reducing the number of trainable parameters while still enabling meaningful task-specific learning. Key hyperparameters, such as the rank of the low-rank matrices, the scaling factor, dropout rate, and the choice of target modules, directly influence both training dynamics and downstream performance. Consequently, understanding and optimizing these parameters is crucial for achieving high-quality results, especially for complex structured-output tasks such as requirement-based test case and test step generation.

## 2.4. Research gaps

Despite the growing body of research on LLMs in software engineering, several important gaps remain. First, there is a lack of systematic comparison across different LLM families for requirement-based test case generation. Existing studies often focus on a single model or compare a limited number of approaches under varying experimental conditions, making it difficult to draw reliable conclusions about relative performance. In particular, there is insufficient empirical evidence comparing open-source models and proprietary models under controlled fine-tuning settings. This limits practitioners' ability to make informed decisions about model selection, especially when considering trade-offs between performance, cost, model size, and deployability.

Second, although parameter-efficient fine-tuning methods such as LoRA are increasingly used, the impact of LoRA hyperparameters on downstream software engineering tasks remains underexplored. Many prior works adopt default or heuristic parameter settings without systematic justification or analysis. As a result, it is unclear how choices such as rank, scaling factor, or target layers influence model behavior and output quality in structured generation tasks like test case and test step generation. Furthermore, automated evaluation methods for generated test artifacts are still insufficiently studied. Most existing evaluations rely on small-scale human assessments or coarse-grained metrics that do not fully capture the quality dimensions of test cases, such as requirement coverage, logical consistency, and practical usability. This lack of robust, scalable evaluation frameworks hinders reproducibility and slows progress in the field, underscoring the need for automated yet reliable assessment approaches.

## 3. Task definition and dataset

### 3.1. Problem formulation

In this work, we formulate automated test case generation as a requirement-to-test transformation task. The goal is to automatically generate complete and structured test artifacts from natural language software requirements using LLMs.

The input to the model consists of:

- Requirement name: a concise textual identifier describing the feature or change request.
- Requirement description: a detailed natural language specification, often including acceptance criteria, functional constraints, and expected system behavior.

Given these inputs, the model is expected to generate the following outputs:

- Test case name: a concise and descriptive title summarizing the intent of the test.
- Test case type: a classification of the test (e.g., functional UI, functional API, regression, usability).
- Test case description: a high-level explanation of the test objective and scope.
- Test steps: a structured, ordered list of executable test steps, typically including actions and expected results.

This formulation requires the model to not only understand the semantics of natural language requirements but also to produce well-structured, domain-specific outputs that align with established software testing practices. The task therefore combines natural language understanding, structured text generation, and domain adaptation.

**3.2. Dataset description**

The dataset used in this study was collected from an industrial software testing environment, consisting of real-world software requirements and their corresponding manually authored test cases. Each data instance links a requirement artifact to one associated test case, ensuring strong traceability between inputs and outputs. The requirements and test cases originate from a test management system and cover a wide range of functional scenarios, including UI behavior, API interactions, integration logic, and configuration management.

The dataset is divided into a training set and a test set. The training set contains 2,583 instances, while the test set consists of 288 instances. Each instance includes a requirement name and description paired with a corresponding test case name, test case type, test case description, and test steps. The requirements are predominantly written in natural language and often include acceptance criteria, numbered conditions, and domain-specific terminology. The test cases reflect practical testing practices and vary in complexity, ranging from short validation checks to long, multi-step procedural tests.

The dataset primarily targets enterprise software systems, with a strong emphasis on functional correctness, user interface validation, API behavior, and system integration. This diversity makes the dataset well-suited for evaluating the ability of LLMs to generalize across different testing contexts and output granular, structured test artifacts.

Before fine-tuning, the dataset underwent standard textual preprocessing steps. These included normalization of whitespace, removal of redundant formatting artifacts, and cleaning of noisy text segments. In addition, special characters, markup tokens, and tool-specific syntax originating from the test automation platform (such as internal identifiers, UI-specific placeholders, and system-generated tokens) were removed when they did not contribute meaningful semantic information for language model training. This step was essential to prevent the model from overfitting to artifacts irrelevant to the downstream generation task.

After preprocessing, each instance was converted into a consistent input–output format suitable for supervised fine-tuning. The resulting dataset provides clean, high-quality training signals for learning the mapping from natural language requirements to structured test cases, while preserving the richness and realism of industrial test documentation.

**3.3. Prompt and output schema design**

To fine-tune LLMs for requirement-based test case generation, we designed a structured, instruction-driven prompt that explicitly defines the task, expected outputs, and domain constraints. The prompt is consistently applied to every training and evaluation sample to ensure uniform learning signals and to minimize ambiguity during both fine-tuning and inference. This design follows an instruction-tuning paradigm, enabling the models to learn how to map natural language requirements to structured and actionable test artifacts.

The prompt first establishes the role of the model as a "test automation assistant" and clearly specifies the input artifacts, namely the requirement name and requirement description. It then enumerates the expected outputs in a fixed order: test case name, test case type, test case description, and test steps. For each output field, the prompt provides detailed guidance on content, level of detail, and formatting. For example, the test case name is required to be concise, descriptive, and traceable to the requirement, while the test case type must be selected from a predefined set of categories. This explicit constraint helps the model produce consistent and standardized outputs that align with real-world testing practices.

A key aspect of the prompt design is the explicit structuring of test steps. The prompt instructs the model to generate step-by-step procedures with clearly ordered actions and corresponding system responses or validations. For API-related scenarios, additional requirements are imposed, such as specifying HTTP methods, endpoints, payloads, headers, and expected response codes. This structured guidance encourages the generation of technically detailed and executable test steps, rather than high-level or purely descriptive text. Furthermore, the prompt instructs the model to cover both standard execution paths and relevant edge cases derived from the acceptance criteria, while avoiding verbatim copying of the requirement text.

The prompt also supports the inclusion of domain knowledge as an optional contextual input. This allows additional system-specific or organizational information to be injected when available, improving adaptability across different projects or environments. Finally, the output schema is explicitly defined at the end of the prompt, listing each required field with its corresponding placeholder. This enforces a consistent output format across all generated samples, which is critical for automated evaluation and downstream processing. Overall, this carefully designed prompt and output schema play a central role in enabling effective fine-tuning and reliable generation of high-quality test cases from natural language requirements. The complete prompt is given in Appendix A.

# 4. Methodology

## 4.1. Overview of the experimental pipeline

This study follows a structured and end-to-end research, development, and experimentation pipeline designed to systematically evaluate LLMs for automated test case generation. The overall pipeline consists of seven main stages: (1) data processing and preparation, where raw requirement–test case pairs are cleaned and transformed into a consistent training format; (2) model selection, covering both open-source and proprietary LLMs; (3) prompt design and engineering, in which a unified instruction-based prompt and output schema are defined; (4) fine-tuning configuration, including the setup of parameter-efficient fine-tuning mechanisms such as LoRA; (5) model fine-tuning, where models are adapted to the downstream task using the prepared dataset; (6) evaluation configuration, including the design of automated evaluation prompts and scoring criteria; and (7) execution of evaluation experiments to assess and compare model performance. This modular pipeline allows individual components to be adjusted or extended while maintaining experimental consistency and reproducibility.

Within this broader workflow, the fine-tuning pipeline itself is composed of four core steps. First, the processed dataset is loaded and prepared for supervised fine-tuning, ensuring correct alignment between inputs and structured outputs. Second, the fine-tuning configuration is defined, focusing on LoRA-specific settings such as target layers and hyperparameters. Third, the LLM is fine-tuned using the configured setup, with the base model weights frozen and only the parameter-efficient components updated. Finally, the fine-tuned model is evaluated using the predefined evaluation framework to measure its ability to generate high-quality test cases and test steps. By clearly separating the fine-tuning process from the surrounding experimental stages, this methodology supports systematic analysis of model behavior, controlled comparison across configurations, and incremental refinement of both models and evaluation strategies.

The source codes, data, experimental configurations, and supplementary materials are available online at: https://github.com/mmoradi-iut/LoRA-LLM-FineTuning.

## 4.2. Models under study

This study evaluates a diverse set of LLMs to investigate the effectiveness of parameter-efficient fine-tuning for automated test case generation. The selected models include both open-source models, which can be deployed and fine-tuned locally, and proprietary models, which are accessed through commercial cloud platforms with built-in fine-tuning capabilities. This combination enables a comprehensive comparison across different model architectures, parameter scales, and deployment paradigms, providing insights into performance-cost trade-offs and practical adoption considerations.

### 4.2.1. Open-source models

To evaluate the performance of locally deployable models under controlled fine-tuning settings, we selected three representative open-source instruction-tuned models with approximately 8 billion parameters or comparable reasoning capacity. These models were chosen due to their strong performance in general language understanding and generation tasks, availability for local fine-tuning, and architectural diversity.

**DeepSeek** [44]

We used a dense model with approximately 1.78 billion parameters, derived from a Qwen-based architecture and distilled from a DeepSeek R1 reasoning model using chain-of-thought data. The model follows a decoder-only dense transformer architecture inherited from Qwen and was further improved through reinforcement learning and knowledge distillation techniques. Despite having fewer parameters than the other models studied, it incorporates reasoning-oriented training signals, which makes it an interesting candidate for structured generation tasks such as test case creation. The model supports a maximum context window of 32,768 tokens, enabling it to process relatively long requirement descriptions and structured prompts.

**Llama** [45]

We employed the Llama-3.1-8B-Instruct model, which contains approximately 8 billion parameters and uses an auto-regressive decoder-only transformer architecture. The model was pre-trained on roughly 15 trillion tokens of multilingual text and code data and subsequently fine-tuned using supervised fine-tuning and Reinforcement Learning from Human Feedback (RLHF) [46]. Llama-3.1 provides strong instruction-following capabilities and supports a large context window of up to 128K tokens, making it suitable for tasks involving long inputs and structured outputs. Its balance between performance and computational efficiency makes it a widely adopted baseline for research and industrial applications.

**Mistral** [47]

We also evaluated the Ministral-8B-Instruct model, which contains approximately 8 billion parameters and is based on a dense decoder-only transformer architecture enhanced with Grouped-Query Attention (GQA) [48] and interleaved sliding-window attention mechanisms. These architectural optimizations improve memory efficiency and long-context handling while maintaining strong generation quality. The model supports a 128K token context window and was pre-trained on multilingual, code-rich web-scale datasets, followed by instruction tuning to improve task-following behavior in conversational and structured generation scenarios. Its efficient architecture and strong performance make it a compelling candidate for parameter-efficient fine-tuning experiments.

### 4.2.2. Proprietary models

In addition to open-source models, we evaluated models from the GPT-4.1 family, accessed through Microsoft Azure's managed fine-tuning infrastructure. These models represent state-of-the-art proprietary LLMs with advanced instruction-following, reasoning, and coding capabilities. The GPT-4.1 family includes three variants with different performance–latency trade-offs:

- GPT-4.1: the full-size flagship model with the highest capability.
- GPT-4.1-mini: a mid-sized variant designed to balance performance and cost.
- GPT-4.1-nano: a lightweight variant optimized for low latency and efficiency.

A key distinguishing feature of the GPT-4.1 family is its extremely large context capacity of up to 1 million tokens, which substantially exceeds that of most open-source models. These models also provide improved instruction adherence, reasoning ability, and code understanding compared to earlier generations such as GPT-4o. Furthermore, the GPT-4.1 family offers improved cost efficiency relative to prior proprietary models, making them more practical for enterprise-scale applications.

By including multiple GPT-4.1 variants alongside open-source models, this study enables a comprehensive comparison between locally fine-tuned models and cloud-based proprietary models, highlighting differences in scalability, performance, and deployment considerations for automated test generation tasks.

### 4.3. Parameter-efficient fine-tuning with LoRA

To efficiently adapt LLMs to the requirement-based test case generation task, this study employs LoRA [42], a widely used parameter-efficient fine-tuning technique. Instead of updating all model parameters during training, LoRA introduces a small number of additional trainable parameters into selected layers of the network while keeping the original pre-trained weights frozen. This approach significantly reduces computational cost, memory consumption, and training time, making it particularly suitable for fine-tuning billion-parameter models on domain-specific tasks with limited hardware resources.

### 4.3.1. LoRA mechanism

The core idea behind LoRA is that the task-specific weight updates required to adapt a pre-trained model often lie in a low-dimensional subspace. Rather than learning full-rank updates for large weight matrices, LoRA decomposes the update into two smaller matrices with reduced rank. Specifically, for a given weight matrix $W$, LoRA introduces two trainable matrices $A$ and $B$ such that the adapted weight becomes:

$$W' = W + \Delta W = W + BA$$

where $A \in \mathbb{R}^{r \times d}$ and $B \in \mathbb{R}^{k \times r}$, and $r$ is the rank hyperparameter with $r \ll \min(d, k)$. During training, only the low-rank matrices $A$ and $B$ are updated, while the original weight matrix $W$ remains unchanged. This dramatically reduces the number of trainable parameters while still enabling the model to learn meaningful task-specific adaptations.

In transformer-based language models, LoRA is typically applied to linear projection layers, particularly those in the attention mechanism (e.g., query, key, value, and output projections) and sometimes feed-forward layers. By injecting low-rank adapters into these components, the model can effectively adjust attention patterns and representation transformations to better align with the downstream task. An additional scaling factor is used to control the magnitude of the low-rank update during training, helping stabilize optimization. Because LoRA modules can be easily attached and removed, multiple task-specific adapters can be maintained without duplicating the full base model, enabling efficient experimentation and deployment.

### 4.3.2. Tuned LoRA parameters

The effectiveness of LoRA depends strongly on the choice of hyperparameters that govern the capacity and behavior of the low-rank adaptation. In this study, we systematically explored several key LoRA parameters to identify configurations that yield optimal performance for automated test case generation.

The first parameter is the rank ($r$), which determines the dimensionality of the low-rank matrices and directly controls the expressive capacity of the adaptation. Higher ranks allow the model to learn more complex task-specific transformations but increase the number of trainable parameters and computational overhead. Lower ranks improve efficiency but may limit adaptation capability. We experimented with multiple rank values to evaluate the trade-off between performance and efficiency.

The second parameter is the scaling factor ($\alpha$), which controls the magnitude of the LoRA update relative to the frozen base weights. The scaling factor effectively rescales the learned low-rank matrices during training and inference, influencing training stability and convergence behavior. Proper tuning of this parameter is important to ensure that the adaptation neither overwhelms the base model nor becomes too weak to capture task-specific patterns.

The third parameter is the LoRA dropout, which applies dropout to the low-rank adaptation during training to improve generalization and reduce overfitting, particularly when training data is limited. Dropout introduces stochastic regularization, encouraging the model to rely on robust patterns rather than memorizing specific training examples.

By systematically varying these parameters across different models, we were able to analyze how LoRA configuration influences downstream performance. This investigation provides insights into the relationship between parameter-efficient adaptation capacity and task-specific generation quality, which is critical for practical deployment of fine-tuned LLMs in software testing environments.

### 4.4. Fine-tuning setup

This subsection describes the training configuration and computational environment used to fine-tune the LLMs in this study. The setup was designed to ensure stable optimization, efficient resource utilization, and reproducibility across multiple experimental runs while accommodating the constraints of parameter-efficient fine-tuning.

**Training configuration**

All open-source models were fine-tuned using supervised instruction tuning with the processed requirement-test case dataset. The training process was conducted for 10 epochs, using a per-device batch size of 2 and gradient accumulation of 16 steps, resulting in an effective batch size of 32 samples per update. Mixed-precision training with FP16 was enabled to reduce memory consumption and accelerate computation. Model evaluation during training was performed periodically every 200 steps to monitor learning progress and detect potential overfitting.

The optimization process used the Adam optimizer with a learning rate of $1\times10^{-6}$ and a weight decay coefficient of 0.01. These conservative hyperparameter choices were selected to stabilize training when adapting large pre-trained models using LoRA while preventing catastrophic forgetting of the base model knowledge. The loss function used during training was the standard autoregressive cross-entropy loss over the generated output tokens.

For parameter-efficient adaptation, LoRA modules were injected into selected transformer layers, and several configurations were explored to evaluate their impact on downstream performance. Specifically, we experimented with LoRA *rank* values of 16 and 32, scaling factors ($α$) of 8, 16, and 32, and *dropout* rates of 0.1 and 0.5. These parameter ranges were chosen to balance adaptation capacity and regularization strength while maintaining computational efficiency. The training dataset consisted of 2,583 samples, while 288 samples were reserved for testing. No additional data augmentation was applied, and samples were shuffled during training to improve generalization.

**Hardware and software environment**

All experiments involving open-source models were executed on Microsoft Azure Machine Learning (Azure ML) infrastructure using a virtual machine equipped with an NVIDIA H100 GPU, 40 CPU cores, 320 GB of system memory, and 128 GB of disk storage. The use of a high-performance GPU enabled efficient fine-tuning of multi-billion-parameter models under mixed-precision settings and gradient accumulation. Proprietary GPT models were fine-tuned using managed cloud-based infrastructure provided through Azure services, which abstracts the underlying hardware configuration while ensuring scalable and reliable training. The GPT-4.1 models were fine-tuned using the default configurations set by the Azure OpenAI model foundry.

The experimental pipeline was implemented using modern deep learning tooling for LLMs, including PyTorch-based training frameworks and libraries supporting transformer architectures and parameter-efficient fine-tuning. This combination of hardware and software infrastructure enabled consistent experimentation across multiple models and LoRA configurations while maintaining reproducibility and efficient resource utilization.

## 5. Automated evaluation framework

Evaluating the quality of automatically generated test cases is inherently challenging due to the complexity and subjectivity involved in assessing software testing artifacts. Unlike many natural language generation tasks, requirement-based test case generation does not always have a single correct output; multiple valid test cases can satisfy the same requirement while differing in structure, wording, or level of detail. To address this challenge and enable large-scale experimentation, we developed an automated evaluation framework that leverages advanced LLMs to assess generated outputs across multiple quality dimensions. This section presents the motivation, methodology, evaluation criteria, and scoring strategy used to systematically compare model performance in our experiments.

## 5.1. Motivation for automated evaluation

A primary challenge in evaluating generated test cases is the absence of a unique ground truth. In practical software testing environments, different testers may produce equally valid test cases for the same requirement, varying in terminology, ordering of steps, or granularity of validation. Traditional evaluation metrics that rely on direct comparison with reference outputs, such as exact match or token-level similarity, are therefore insufficient for capturing the true quality of generated test artifacts. Effective evaluation must instead consider semantic correctness, requirement coverage, clarity, and adherence to testing best practices, which are difficult to measure using conventional automated metrics.

Another major limitation is the reliance on manual human evaluation, which is time-consuming, expensive, and difficult to scale. Human assessment requires domain expertise and careful review of each generated test case against the requirement specification, making it impractical for large experimental studies involving multiple models and fine-tuning configurations. Manual evaluation also introduces potential inconsistencies due to subjective judgment across evaluators. These challenges motivate the need for a scalable, consistent, and reproducible evaluation approach.

Recent advances in LLMs provide an opportunity to address these issues by using LLMs themselves as evaluators. Modern models demonstrate strong capabilities in reasoning, semantic comparison, and structured judgment tasks, enabling them to assess generated outputs against requirements using predefined criteria. By leveraging these capabilities, automated evaluation can approximate expert judgment while significantly reducing cost and effort. Consequently, this work adopts an LLM-based evaluation strategy to support comprehensive and efficient comparison of fine-tuned models for automated test case generation.

## 5.2. GPT-4o-based evaluation method

To enable scalable and consistent evaluation across multiple models and configurations, we adopted a LLM-based evaluation approach, using GPT-4o as an automated judge. GPT-4o was selected due to its strong reasoning ability, instruction-following performance, and demonstrated effectiveness in comparative evaluation tasks involving structured outputs. In this framework, GPT-4o receives the requirement, the model-generated test case, and the corresponding human-authored reference test case, and produces structured evaluation scores across predefined criteria.

The evaluation process is implemented through a deterministic chat-completion call with temperature set to 0 and top-p equal to 1, ensuring consistent scoring behavior across repeated runs. The model is instructed through a system prompt that defines its role as an expert automated evaluator in software testing, followed by a user prompt containing detailed evaluation instructions and the input artifacts. The evaluation prompt explicitly defines the context, scoring scale, criteria definitions, and required output format. By constraining the output to a JSON-style dictionary containing numeric scores, the framework supports automated parsing and downstream aggregation without manual intervention. This design ensures reproducibility, minimizes response variability, and enables efficient evaluation of large numbers of generated test cases.

Importantly, GPT-4o is not used to replace human judgment entirely but rather to approximate expert-level assessment at scale. Prior research has shown that strong LLMs can serve as reliable evaluators when provided with well-defined criteria and structured prompts [49-52]. In our context, the model's ability to reason about semantics, structure, and technical correctness makes it particularly suitable for evaluating requirement-based test artifacts.

### 5.3. Evaluation criteria

The evaluation framework assesses generated test cases using nine complementary criteria, designed to capture multiple dimensions of quality relevant to software testing. These criteria extend beyond simple textual similarity and focus on semantic correctness, completeness, clarity, and practical usability.

- **Semantic Similarity** evaluates how closely the overall meaning and intent of the generated test case align with the reference test case, emphasizing conceptual alignment rather than lexical overlap.
- **Information Coverage** measures how comprehensively the generated output captures important details such as preconditions, actions, expected outcomes, and edge cases.
- **Critical Content Match** assesses whether essential elements such as roles, system components, data attributes, or UI elements are correctly preserved.
- **Structural and Format Accuracy** evaluates whether the output follows the expected structure, including a clear title, type classification, and logically ordered test steps.
- **Omission** measures the extent to which important information is missing, with higher scores indicating minimal missing content.
- **Hallucination** assesses whether the generated test case introduces unsupported or irrelevant information not grounded in the requirement or reference.
- **Ambiguity** evaluates the clarity and precision of the language used, focusing on whether the test case can be easily understood and executed.
- **Redundancy** measures the degree of unnecessary repetition, with higher scores indicating concise and well-organized outputs.

- **Diversity and Novelty** captures whether the generated test case introduces valid alternative interpretations or improvements beyond the reference while remaining logically sound.

Each criterion is scored on a 0–100 ordinal scale, where 0 indicates completely missing or incorrect content and 100 indicates excellent quality. This multi-dimensional evaluation provides a more comprehensive assessment than single-metric approaches and allows fine-grained analysis of model strengths and weaknesses. The prompt used for the automated evaluation is presented in Appendix B.

## 5.4. Scoring and aggregation strategy

For each generated test case, GPT-4o produces a dictionary containing nine scores corresponding to the evaluation criteria. These scores are then aggregated to compute overall performance metrics for individual samples, models, and experimental configurations. Specifically, the per-sample score is calculated as the arithmetic mean of the nine criterion scores, providing a normalized quality measure between 0 and 100. This averaging approach ensures that no single criterion dominates the evaluation while still reflecting overall generation quality.

At the model level, performance is summarized by computing the mean score across all test samples in the evaluation dataset. Additional descriptive statistics, such as standard deviation, can be used to analyze variability across samples. For comparative experiments, aggregated scores enable direct ranking of models and fine-tuning configurations, supporting systematic analysis of the impact of LoRA parameters and model architecture choices.

The use of a structured scoring framework also facilitates dimension-level analysis, allowing us to examine how models perform across individual quality aspects such as coverage, hallucination, or structural accuracy. This provides deeper insight into model behavior beyond overall scores and helps identify strengths and weaknesses associated with specific architectures or training configurations. Overall, the combination of deterministic evaluation, multi-criteria scoring, and structured aggregation enables robust and reproducible comparison of automated test generation approaches.

## 6. Results and discussion

Table 1 presents the evaluation results of the DeepSeek model before and after applying LoRA-based fine-tuning across different configurations of *rank*, scaling factor (*α*), and *dropout*.

The baseline performance of the model before fine-tuning shows an overall score of 62.85, indicating moderate capability in generating requirement-based test cases. The highest baseline scores are observed in structural and format accuracy (80.23), ambiguity (80.75), and redundancy (82.44), suggesting that the model is already capable of producing relatively well-structured and readable outputs. However, the model performs less effectively in criteria that require deeper understanding of the requirement and preservation of detailed information, such as information coverage (50.28), critical content match (52.15), and diversity and novelty (43.67). These results highlight that, without task-specific adaptation, the model struggles to fully capture the semantics and detailed content necessary for comprehensive test case generation.

**Table 1.** Performance scores obtained by the DeepSeek model before and after fine-tuning on the test case generation dataset across different LoRA configurations. The highest score obtained for each evaluation criterion is shown in underlined format.

| LoRA α | 8 | | | | 16 | | | | 32 | | | | Before fine-tuning |
|---|---|---|---|---|---|---|---|---|---|---|---|---|---|
| LoRA rank | 16 | | 32 | | 16 | | 32 | | 16 | | 32 | | |
| LoRA dropout | 0.1 | 0.5 | 0.1 | 0.5 | 0.1 | 0.5 | 0.1 | 0.5 | 0.1 | 0.5 | 0.1 | 0.5 | |
| Semantic similarity | 62.94 | 63.71 | 64.10 | <u>65.32</u> | 57.49 | 58.30 | 60.94 | 62.57 | 55.13 | 55.71 | 56.49 | 57.16 | **60.19** |
| Information coverage | 52.18 | 54.51 | 53.73 | <u>54.78</u> | 47.13 | 48.54 | 51.40 | 52.68 | 46.75 | 47.28 | 46.22 | 48.81 | **50.28** |
| Critical content match | 53.70 | 53.95 | 54.39 | <u>55.06</u> | 48.54 | 51.47 | 52.71 | 54.36 | 49.38 | 49.93 | 48.07 | 50.35 | **52.15** |
| Structural and format accuracy | 81.84 | 83.16 | 84.07 | <u>84.51</u> | 76.82 | 76.05 | 78.30 | 80.91 | 74.21 | 75.19 | 74.00 | 75.83 | **80.23** |
| Omission | 49.65 | 51.28 | 52.68 | <u>52.90</u> | 47.08 | 47.68 | 48.59 | 50.17 | 42.97 | 44.10 | 43.53 | 46.55 | **47.31** |
| Hallucination | 70.37 | 72.03 | 73.14 | <u>74.16</u> | 68.17 | 67.51 | 69.08 | 69.83 | 65.06 | 65.85 | 66.51 | 67.80 | **68.70** |
| Ambiguity | 81.46 | <u>82.64</u> | 82.18 | 81.93 | 77.31 | 78.92 | 80.52 | 82.05 | 75.12 | 76.33 | 75.69 | 78.34 | **80.75** |
| Redundancy | 84.20 | 83.82 | <u>84.55</u> | 83.49 | 81.19 | 81.70 | 83.23 | 83.94 | 79.03 | 79.62 | 80.46 | 80.04 | **82.44** |
| Diversity and novelty | 44.51 | 47.19 | 48.53 | <u>49.35</u> | 42.70 | 44.16 | 45.64 | 46.72 | 40.76 | 41.47 | 41.18 | 43.26 | **43.67** |
| Overall | 64.53 | 65.81 | 66.37 | <u>66.82</u> | 60.71 | 61.59 | 63.37 | 64.80 | 58.71 | 59.49 | 59.12 | 60.90 | **62.85** |

After applying LoRA-based fine-tuning, consistent performance improvements can be observed across most evaluation metrics and configurations. The best results are achieved when $α=8$ and $rank=32$ with $dropout=0.5$, producing the highest overall score of 66.82, which represents a substantial improvement over the baseline. In this configuration, improvements are particularly evident in semantic similarity (65.32), information coverage (54.78), and critical content match (55.06), indicating that the fine-tuned model becomes better at understanding requirements and preserving essential test-related information. Additionally, structural qualities such as format accuracy, hallucination control, and ambiguity remain consistently strong, demonstrating that fine-tuning enhances semantic alignment without degrading structural quality. Across all configurations, higher LoRA ranks generally yield better performance, suggesting that increased adaptation capacity improves the model's ability to learn task-specific patterns. However, performance declines slightly for higher $α$ values (16 and 32), indicating that excessively large scaling factors may lead to less stable adaptation. Overall, these results confirm that LoRA-based fine-tuning significantly improves the DeepSeek model's ability to generate accurate and comprehensive test cases while maintaining structural quality.

Table 2 shows that the Llama-3.1-8B-Instruct model achieves stronger performance than DeepSeek both before and after fine-tuning. Before fine-tuning, Llama obtains an overall score of 63.77, which is higher than the DeepSeek baseline of 62.85. Similar to DeepSeek, Llama is strongest in structural and format accuracy (81.20), ambiguity (82.17), and redundancy (85.22), indicating that even the untuned model can produce outputs that are relatively well organized, readable, and non-repetitive. At the same time, lower scores in information coverage (47.88), omission (48.54), and diversity and novelty (45.75) show that the model still misses important requirement details and tends to produce limited variations in test design without task-specific adaptation. Compared with DeepSeek, however, Llama starts from a stronger baseline in several important criteria, including semantic similarity, critical content match, and overall score, suggesting a better natural fit for requirement-to-test-case generation even prior to fine-tuning.

**Table 2.** Performance scores obtained by the Llama model before and after fine-tuning on the test case generation dataset across different LoRA configurations. The highest score obtained for each evaluation criterion is shown in underlined format.

| LoRA α | 8 | | | | 16 | | | | 32 | | | | Before fine-tuning |
|---|---|---|---|---|---|---|---|---|---|---|---|---|---|
| LoRA rank | 16 | | 32 | | 16 | | 32 | | 16 | | 32 | | |
| LoRA dropout | 0.1 | 0.5 | 0.1 | 0.5 | 0.1 | 0.5 | 0.1 | 0.5 | 0.1 | 0.5 | 0.1 | 0.5 | |
| Semantic similarity | 65.11 | <u>67.80</u> | 66.38 | 67.25 | 61.07 | 62.49 | 64.70 | 65.62 | 58.06 | 58.84 | 60.15 | 62.72 | **61.53** |
| Information coverage | 50.75 | 52.44 | 51.07 | <u>53.37</u> | 47.14 | 47.80 | 50.28 | 51.03 | 45.17 | 45.61 | 46.54 | 47.39 | **47.88** |
| Critical content match | 57.02 | <u>58.19</u> | 57.55 | 57.32 | 54.31 | 55.74 | 56.47 | 57.30 | 51.80 | 51.27 | 53.25 | 55.28 | **53.62** |
| Structural and format accuracy | 84.18 | 84.60 | 85.10 | 84.87 | 80.96 | 81.45 | <u>85.72</u> | 83.91 | 77.32 | 78.13 | 80.47 | 82.93 | **81.20** |
| Omission | 51.36 | <u>52.75</u> | 52.41 | 51.74 | 48.53 | 50.06 | 51.09 | 51.70 | 45.91 | 46.30 | 47.61 | 49.54 | **48.54** |
| Hallucination | 71.88 | <u>74.08</u> | 72.20 | 73.13 | 68.75 | 70.38 | 72.15 | 72.53 | 67.44 | 67.02 | 69.26 | 69.90 | **68.06** |
| Ambiguity | 84.25 | <u>85.19</u> | 83.79 | 84.49 | 83.80 | 84.51 | 83.63 | 84.96 | 78.63 | 79.08 | 81.30 | 84.05 | **82.17** |
| Redundancy | 85.64 | 86.04 | 85.03 | 85.60 | 84.23 | 85.05 | 84.95 | <u>86.44</u> | 80.12 | 81.19 | 83.00 | 85.73 | **85.22** |
| Diversity and novelty | 48.78 | <u>50.33</u> | 48.20 | 49.51 | 45.79 | 47.19 | 48.31 | 49.19 | 41.05 | 41.77 | 43.87 | 47.41 | **45.75** |
| Overall | 66.55 | <u>67.93</u> | 66.85 | 67.47 | 63.84 | 64.96 | 66.36 | 66.96 | 60.61 | 61.02 | 62.82 | 64.99 | **63.77** |

After LoRA-based fine-tuning, Llama shows clear and consistent improvements across nearly all evaluation criteria. The best overall result is obtained with LoRA *α*=8, *rank*=16, and *dropout*=0.5, reaching an overall score of 67.93, which exceeds the best DeepSeek result of 66.82. Under this best configuration, Llama achieves particularly strong scores in semantic similarity (67.80), hallucination (74.08), ambiguity (85.19), and redundancy (86.04), indicating that the fine-tuned model not only aligns more closely with the reference test cases but also produces clearer and more reliable outputs. As with DeepSeek, the results suggest that lower *α* values tend to work better, while increasing α to 16 or 32 generally leads to some degradation in performance. Unlike DeepSeek, however, Llama maintains comparatively strong results across a broader range of configurations, which may indicate greater robustness to LoRA hyperparameter changes. Overall, the results demonstrate that Llama benefits substantially from LoRA-based fine-tuning and outperforms DeepSeek in both baseline and fine-tuned settings, making it a stronger open-source candidate for automated test case generation.

Table 3 shows that the Ministral-8B-Instruct model achieves the strongest results among the open-source models, both before and after fine-tuning. Before fine-tuning, it reaches an overall score of 65.88, clearly outperforming the corresponding baselines of DeepSeek (62.85) and Llama (63.77). This advantage is visible across nearly all evaluation criteria, especially structural and format accuracy (84.70), hallucination (71.26), ambiguity (84.09), and redundancy (89.72), indicating that the model already produces highly structured, clear, and reliable test cases even without task-specific adaptation. In addition, its baseline scores in semantic similarity (64.31), information coverage (49.75), and critical content match (55.13) are competitive and generally stronger than those of DeepSeek and comparable to or better than those of Llama. These results suggest that Mistral has a stronger initial capability for transforming requirements into coherent and practically useful test cases.

**Table 3.** Performance scores obtained by the Mistral model before and after fine-tuning on the test case generation dataset across different LoRA configurations. The highest score obtained for each evaluation criterion is shown in underlined format.

| LoRA α | 8 | | | | 16 | | | | 32 | | | | Before fine-tuning |
|---|---|---|---|---|---|---|---|---|---|---|---|---|---|
| LoRA rank | 16 | | 32 | | 16 | | 32 | | 16 | | 32 | | |
| LoRA dropout | 0.1 | 0.5 | 0.1 | 0.5 | 0.1 | 0.5 | 0.1 | 0.5 | 0.1 | 0.5 | 0.1 | 0.5 | |
| Semantic similarity | 68.06 | 69.17 | 66.43 | 66.15 | 66.94 | 67.61 | 68.21 | 68.21 | 59.07 | 61.45 | 64.82 | 66.27 | **64.31** |
| Information coverage | 53.82 | 55.30 | 51.30 | 51.73 | 51.02 | 51.17 | 52.87 | 53.19 | 43.62 | 46.73 | 48.33 | 50.66 | **49.75** |
| Critical content match | 60.17 | 61.08 | 57.53 | 57.08 | 58.11 | 58.70 | 60.04 | 60.73 | 51.30 | 52.81 | 54.67 | 56.74 | **55.13** |
| Structural and format accuracy | 88.78 | 90.27 | 85.96 | 86.53 | 86.21 | 86.53 | 87.69 | 88.15 | 79.44 | 82.39 | 84.05 | 86.12 | **84.70** |
| Omission | 53.30 | 54.42 | 51.67 | 50.62 | 51.87 | 52.40 | 53.17 | 53.08 | 44.19 | 46.15 | 47.91 | 50.03 | **48.83** |
| Hallucination | 76.22 | 77.08 | 73.80 | 73.29 | 74.75 | 75.02 | 75.43 | 75.90 | 65.72 | 69.56 | 70.18 | 72.48 | **71.26** |
| Ambiguity | 87.61 | 89.02 | 85.81 | 86.25 | 86.83 | 86.65 | 87.01 | 87.22 | 78.35 | 81.10 | 85.52 | 86.01 | **84.09** |
| Redundancy | 93.25 | 94.09 | 91.46 | 91.46 | 91.95 | 92.21 | 92.42 | 92.75 | 81.90 | 85.11 | 87.83 | 91.07 | **89.72** |
| Diversity and novelty | 50.34 | 51.63 | 46.75 | 47.07 | 48.38 | 48.79 | 50.14 | 51.04 | 42.18 | 42.65 | 43.57 | 46.25 | **45.13** |
| Overall | 70.17 | 71.34 | 67.85 | 67.79 | 68.45 | 68.78 | 69.66 | 70.03 | 60.64 | 63.10 | 65.20 | 67.29 | **65.88** |

After LoRA-based fine-tuning, the Mistral model improves further and delivers the best open-source performance observed in this study. The highest overall score, 71.34, is obtained with LoRA α=8, rank=16, and dropout=0.5, which is notably higher than the best fine-tuned results of Llama (67.93) and DeepSeek (66.82). Under this configuration, the model achieves very strong scores in semantic similarity (69.17), information coverage (55.30), critical content match (61.08), structural and format accuracy (90.27), hallucination (77.08), ambiguity (89.02), and redundancy (94.09). These results indicate that fine-tuning substantially strengthens both the semantic alignment and the structural quality of the generated test cases. Similar to the previous models, the best results are generally achieved with lower $α$ values, while larger $α$ values, especially $α=32$, tend to reduce performance relative to the best settings. However, even under less favorable configurations, Mistral still remains highly competitive, showing greater robustness than DeepSeek and consistently outperforming Llama across most criteria. Overall, the results confirm that Mistral is the most effective open-source model for this task, and that LoRA-based fine-tuning can raise its performance to a level substantially above the other open-source alternatives.

Table 4 presents the performance of the GPT-4.1 family models before and after fine-tuning, including the flagship GPT-4.1, the mid-sized GPT-4.1-mini, and the lightweight GPT-4.1-nano. Before fine-tuning, all three models already demonstrate strong performance, with overall scores of 74.15, 74.10, and 72.76, respectively. These results significantly exceed those of all open-source models, highlighting the strong baseline capabilities of proprietary models in requirement-based test case generation. In particular, the models achieve very high scores in structural and format accuracy (above 91), ambiguity (above 90), and redundancy (above 93), indicating that they consistently produce well-structured, clear, and concise test cases. Among the three variants, GPT-4.1 and GPT-4.1-mini perform almost identically, with GPT-4.1-mini slightly outperforming GPT-4.1 in several criteria such as semantic similarity and information coverage, suggesting that the smaller model retains strong reasoning and generation capabilities despite reduced size.

After fine-tuning, further improvements are observed for GPT-4.1 and GPT-4.1-mini, with overall scores increasing to 78.32 and 79.74, respectively. These gains are particularly evident in information coverage, critical content match, and diversity and novelty, indicating that fine-tuning enhances the models' ability to capture detailed requirement information and generate more comprehensive and varied test cases. GPT-4.1-mini achieves the best overall performance among all evaluated models, demonstrating that a balanced model size can provide an optimal trade-off between capability and adaptability. In contrast, GPT-4.1-nano shows a notable decline after fine-tuning, with its overall score dropping to 64.71, primarily due to significant decreases in structural accuracy, ambiguity, and redundancy. This suggests that smaller models may be more sensitive to fine-tuning and can suffer from instability or reduced generalization when adapted to complex structured generation tasks. Overall, these results confirm that fine-tuning further enhances the already strong performance of larger GPT-4.1 models, while also highlighting important differences in robustness and adaptability across model sizes within the same family.

Table 4. Performance scores obtained by the GPT-4.1 models before and after fine-tuning on the test case generation dataset. The highest score obtained for each evaluation criterion is shown in underlined format.

| | Before fine-tuning | | | After fine-tuning | | |
|---|---|---|---|---|---|---|
| | GPT-4.1 | GPT-4.1-mini | GPT-4.1-nano | GPT-4.1 | GPT-4.1-mini | GPT-4.1-nano |
| Semantic similarity | 70.62 | 70.83 | 70.48 | 72.77 | <u>73.81</u> | 66.94 |
| Information coverage | 58.61 | 59.65 | 56.45 | 66.73 | <u>68.95</u> | 53.81 |
| Critical content match | 65.83 | 66.25 | 63.12 | 70.06 | <u>72.36</u> | 58.47 |
| Structural and format accuracy | 93.40 | 92.36 | 91.94 | <u>97.73</u> | 97.29 | 83.61 |
| Omission | 58.19 | 58.88 | 55.97 | 65.62 | <u>68.61</u> | 52.84 |
| Hallucination | 80.27 | 80.55 | 79.51 | 83.05 | <u>84.16</u> | 64.93 |
| Ambiguity | 92.15 | 91.38 | 90.69 | <u>95.0</u> | <u>95.0</u> | 80.48 |
| Redundancy | 95.97 | 94.44 | 93.61 | <u>94.23</u> | 93.61 | 72.56 |
| Diversity and novelty | 52.36 | 52.56 | 53.05 | 60.69 | <u>63.88</u> | 48.75 |
| Overall | 74.15 | 74.10 | 72.76 | 78.32 | <u>79.74</u> | 64.71 |

When comparing all evaluated models, a clear performance hierarchy emerges between the open-source and proprietary approaches. Among the open-source models, Ministral-8B-Instruct consistently achieves the best results, outperforming both Llama-3.1-8B and DeepSeek across most evaluation metrics and configurations. Its strong baseline performance, combined with substantial gains after LoRA-based fine-tuning, demonstrates its superior capability in capturing requirement semantics, preserving critical content, and producing well-structured test cases. In contrast, while Llama shows solid and robust improvements after fine-tuning, and DeepSeek benefits significantly from adaptation despite its smaller size, both models remain below Mistral in overall performance. However, all open-source models exhibit a common trend: fine-tuning leads to meaningful improvements, particularly in semantic similarity, information coverage, and critical content match, confirming the effectiveness of parameter-efficient adaptation for this task.

Despite these improvements, the GPT-4.1 family models still achieve the highest overall performance, especially after fine-tuning, with GPT-4.1-mini delivering the best results across all evaluated models. These models demonstrate superior capabilities in both semantic understanding and structured generation, as reflected in consistently high scores across all evaluation criteria. However, the performance gap between proprietary and open-source models narrows significantly when considering the best fine-tuned open-source model. Notably, the fine-tuned Mistral model achieves performance comparable to the pre-fine-tuned GPT-4.1 models, which is a highly valuable result given its significantly smaller size and lower operational cost. This highlights a key practical advantage of open-source models: they offer greater flexibility, cost efficiency, and deployability, allowing organizations to run models locally, customize them extensively, and avoid dependency on external APIs. In contrast, proprietary models provide higher out-of-the-box performance and ease of use but come with trade-offs in terms of cost, limited control over fine-tuning processes, and potential constraints related to data privacy and deployment. Therefore, the choice between open-source and proprietary models depends on application requirements, with open-source models being particularly attractive for scenarios where cost, customization, and data control are critical, while proprietary models remain preferable for maximum performance and minimal setup effort.

# 7. Conclusion

This paper addressed the challenge of automated test case generation from natural language requirements, a task that remains highly complex due to the ambiguity of requirements and the structured nature of test artifacts. Motivated by the limitations of rule-based and traditional machine learning approaches, and the growing potential of LLMs, we investigated how LLMs can be effectively adapted for this task. In particular, we focused on parameter-efficient fine-tuning using LoRA, aiming to balance performance with computational efficiency. To support a comprehensive and scalable evaluation, we also introduced an automated evaluation framework based on GPT-4o, enabling consistent assessment across multiple models and configurations.

From a methodological perspective, we conducted a large-scale empirical study involving multiple open-source and proprietary models, including DeepSeek, Llama, Mistral, and GPT-4.1 variants. We systematically explored different LoRA configurations, varying rank, scaling factor, and dropout, and evaluated their impact using a multi-criteria evaluation framework. The results demonstrate that LoRA-based fine-tuning significantly improves model performance across all open-source models, particularly in semantic understanding, information coverage, and critical content preservation. Among the open-source models, Ministral-8B-Instruct achieved the best performance, showing strong robustness and substantial gains after fine-tuning. Furthermore, the experiments confirm that LoRA hyperparameters have a meaningful effect on downstream performance, highlighting the importance of careful configuration in parameter-efficient fine-tuning.

In comparison, the GPT-4.1 family models achieved the highest overall performance, especially after fine-tuning, demonstrating superior capabilities in both reasoning and structured generation. However, one of the most important findings of this study is that a fine-tuned 8B open-source model can reach performance comparable to pre-fine-tuned GPT-4 models, which has significant practical implications. This result suggests that organizations can achieve competitive performance using smaller, more cost-efficient, and locally deployable models when combined with effective fine-tuning strategies. Overall, this work advances the understanding of LLM-based test automation and provides actionable insights for both researchers and practitioners, particularly in selecting models, configuring fine-tuning approaches, and designing scalable evaluation methods for real-world software testing applications.

**Declaration of generative AI and AI-assisted technologies in the manuscript preparation process**

During the preparation of this work the authors used GPT-5.3 in order to improve language and readability. After using this tool, the authors reviewed and edited the content as needed and take full responsibility for the content of the published article.

[10]  A. Singh, "Taxonomy of Machine Learning Techniques in Test Case Generation," in *2023 7th International Conference on Intelligent Computing and Control Systems (ICICCS)*, 2023, pp. 474-481.

[11]  A. Fan, B. Gokkaya, M. Harman, M. Lyubarskiy, S. Sengupta, S. Yoo*, et al.*, "Large Language Models for Software Engineering: Survey and Open Problems," in *2023 IEEE/ACM International Conference on Software Engineering: Future of Software Engineering (ICSE-FoSE)*, 2023, pp. 31-53.

[12]  X. Hou, Y. Zhao, Y. Liu, Z. Yang, K. Wang, L. Li*, et al.*, "Large Language Models for Software Engineering: A Systematic Literature Review," *ACM Trans. Softw. Eng. Methodol.,* vol. 33, p. Article 220, 2024.

[13]  W. Wang, C. Yang, Z. Wang, Y. Huang, Z. Chu, D. Song*, et al.*, "TestEval: Benchmarking Large Language Models for Test Case Generation," Albuquerque, New Mexico, 2025, pp. 3547-3562.

[14]  M. Schäfer, S. Nadi, A. Eghbali, and F. Tip, "An Empirical Evaluation of Using Large Language Models for Automated Unit Test Generation," *IEEE Transactions on Software Engineering,* vol. 50, pp. 85-105, 2024.

[15]  A. M. Dakhel, A. Nikanjam, V. Majdinasab, F. Khomh, and M. C. Desmarais, "Effective test generation using pre-trained Large Language Models and mutation testing," *Information and Software Technology,* vol. 171, p. 107468, 2024.

[16]  M. Lafi, T. Alrawashed, and A. M. Hammad, "Automated Test Cases Generation From Requirements Specification," in *2021 International Conference on Information Technology (ICIT)*, 2021, pp. 852-857.

[17]  J. Navarro and R. Ibarra, "Automatic test case generation using natural language processing: A systematic mapping study," *Information and Software Technology,* vol. 189, p. 107929, 2026.

[18]  F. Arooj, H. Alishba, and R. Summair, "Automated Test Case Generation From Natural Language Requirements Using Natural Language Processing," *Journal of Computing & Biomedical Informatics,* vol. 9, 09/01 2025.

[19]  H. Ayenew and M. Wagaw, "Software test case generation using natural language processing (NLP): a systematic literature review," *Artificial Intelligence Evolution,* pp. 1-10, 2024.

[20]  Y. Xia, J. Kim, Y. Chen, H. Ye, S. Kundu, C. C. Hao*, et al.*, "Understanding the Performance and Estimating the Cost of LLM Fine-Tuning," in *2024 IEEE International Symposium on Workload Characterization (IISWC)*, 2024, pp. 210-223.

[21]  N. Ding, Y. Qin, G. Yang, F. Wei, Z. Yang, Y. Su*, et al.*, "Parameter-efficient fine-tuning of large-scale pre-trained language models," *Nature Machine Intelligence,* vol. 5, pp. 220-235, 2023.

[22]  Z. Yang, R. Huang, C. Cui, N. Niu, and D. Towey, "Requirements-based test generation: A comprehensive survey," *ACM Transactions on Software Engineering and Methodology,* 2025.

[23]  S. Alagarsamy, C. Tantithamthavorn, W. Takerngsaksiri, C. Arora, and A. Aleti, "Enhancing large language models for text-to-testcase generation," *Journal of Systems and Software,* vol. 230, p. 112531, 2025.

[24]  L. H. Tahat, B. Vaysburg, B. Korel, and A. J. Bader, "Requirement-based automated black-box test generation," in *25th Annual International Computer Software and Applications Conference. COMPSAC 2001*, 2001, pp. 489-495.

[25]  S. Hesari, R. Behjati, and T. Yue, "Towards a systematic requirement-based test generation framework: Industrial challenges and needs," in *2013 21st IEEE International Requirements Engineering Conference (RE)*, 2013, pp. 261-266.
29

**Appendix A**

The prompt used for generating test cases from test requirements. This prompt was employed during fine-tuning the LLMs, also during inference when we evaluated the LLMs.

```
### Instruction:
You are a test automation assistant. Your task is to generate a test
case from a given software requirement. Each requirement includes a name
and a detailed description. Based on this input and given the domain
knowledge, you must generate the following:

    ###Test Case Name:
   - Provide a concise, descriptive name for the test case.
   - It should summarize the module or feature being tested and the
behavior or condition under test.
   - It should be unique, specific, and traceable to the requirement.

   ###Test Case Type:
   Indicate the type of test, selected from the following:
   - Functional - UI (tests user interface behavior)
   - Functional - API (tests backend API behavior)
   - Usability (tests user experience or workflow)
   - Regression (ensures existing functionality is not broken)

   ###Test Case Description:
   - Describe the purpose of the test case.
   - Include what the test validates and any relevant context.
   - Mention key functionality, preconditions, or variations.

   ###Test Steps:
   - Write a step-by-step procedure to perform the test.
   - Each step should be clear, specific, and ordered.
   - Use the structure:
          1. [Setup or action]
          2. [System response or validation]
   - If API testing is involved, include:
```

- HTTP method and endpoint

- Payload and headers (if applicable)

- Expected response and status code

- Any conditions, edge cases, or validation rules

Assume that the system includes both a web-based UI and RESTful APIs. Requirements may involve templates, configuration fields, user permissions, workflows, or API interactions. Acceptance criteria define both standard (happy path) and boundary (error or edge) cases. The generated test case should cover the main path and validate expected system behavior precisely.

Do not copy the requirement verbatim. Instead, interpret it to create an accurate, testable, and technically detailed test case.

Here is the input:

---

Requirement Name:

{requirement_name}

Requirement Description:

{requirement_description}

---

###Domain knowledge

{domain_knowledge}

### Response:

- Test case name: {test_case_name}

- Test case type: {test_case_type}

- Test case description: {test_case_description}

- Test steps: {test_steps}

**Appendix B**

The prompt used for automated evaluation of response generated by the LLMs for the test case generation task.

```
You are acting as an *expert automated evaluator* for test case
generation in a software testing task.

   == CONTEXT ==
   You are given:
   - A software test requirement, which includes a name and description.
   - A model-generated test case, which includes a test case name, type,
description, and test steps.
   - The ground truth (reference) test case created by a human.

   == TASK ==
   You must evaluate the quality of the model-generated test case based
on the following nine advanced criteria. For each, assign a score between
0 and 100, where:
   - 90-100 = Excellent
   - 70-89 = Good
   - 50-69 = Fair
   - 30-49 = Poor
   - 1-29 = Very poor
   - 0 = Completely missing, unrelated, or misleading

   Scores can be any integer value in the range [0, 100], allowing
fine-grained evaluation.

   == EVALUATION CRITERIA ==

   1. Semantic Similarity: How closely the meaning of the model-
generated test case matches the ground truth. Focus on *overall intent*
and key concepts, not literal wording.

   2. Information Coverage: How completely the generated output
includes all the key details, preconditions, actions, expected outcomes,
and edge cases present in the ground truth.
```

3. Critical Content Match: Whether the model has preserved *must-have* elements (e.g., specific actions, roles, UI elements, data types) from the requirement or ground truth.

4. Structural and Format Accuracy: Whether the output is well-structured and conforms to expected formatting:

- Clear test case title and type

- Actionable, step-by-step instructions

- Coherent order of steps

- Consistent formatting

5. Omission: Degree to which important information is *missing*. Higher scores mean minimal omissions.

6. Hallucination: Degree to which the model *invents irrelevant or unsupported content* not present in the requirement. Higher scores mean fewer or no hallucinations.

7. Ambiguity: Clarity and precision of language used. Higher scores mean the test case is easy to understand and unambiguous.

8. Redundancy: Whether the test case contains unnecessary repetition. Higher scores mean the output is concise and avoids redundant content.

9. Diversity and Novelty: If applicable, does the generated test case introduce valid, logically sound variations or interpretations of the requirement that differ meaningfully from the ground truth?

== INPUT ==

### Requirement
- Name: {requirement_name}
- Description: {requirement_description}

### Model-Generated Test Case
{model_response}

### Ground Truth Test Case

- Name: {test_case_name}

- Type: {test_case_type}

- Description: {test_case_description}

- Steps: {test_steps}

== OUTPUT FORMAT ==

Return a JSON-style dictionary with each metric as a key and a score (0-100) as the value.

Example format:

{{

  "semantic_similarity": 92,

  "information_coverage": 78,

  "critical_content_match": 85,

  "structural_format_accuracy": 88,

  "omission": 90,

  "hallucination": 95,

  "ambiguity": 93,

  "redundancy": 87,

  "diversity_novelty": 72

}}

Do not include any explanation, comments, or extra formatting. Only return the dictionary.

"""